\documentclass[12pt]{article}
\usepackage{amssymb}
\usepackage{array}
\newcommand{\be}{\begin{equation}}
\newcommand{\ee}{\end{equation}}
\newcommand{\bea}{\begin{eqnarray}}
\newcommand{\eea}{\end{eqnarray}}
\begin{document}

\title{NON-STATISTICAL $\gamma$ RAYS FROM FRAGMENTS}

\author{F.F.Karpeshin\\
\em  St. Petersburg University, Institute of Physics\\ 
SU-198504 St. Petersburg, Russia	}

\maketitle

\section{Introduction}

Observation of the shake effects brought about by the neck rupture 
in fission is of great interest.
From mathematical view point, the rupture means break-down of the analyticity of the Hamiltonian
with respect to time. Thus, muon shake in muon-induced
prompt fission manifests itself in muonic conversion. The calculated probability agrees
with the experiment \cite{D1,D2}. Herein we calculate the probability of emission of $\gamma$ quanta,
the results are also of interest in connection with experiments \cite{D3,D4}, in which
works non-statistical $\gamma$ rays from $^{252}$Cf spontaneous fission are under investigation.

	The process of snapping-back of the nuclear surface gives rise to the oscillations of the surface,
whose lifetime is determined by the dissipation. It can be evaluated as $\tau_{diss}\approx 10^{-19}$~s~\cite{D5}.
The oscillations generate nonstationary electromagnetic field in space. That causes electromagnetic
processes of internal conversion and $\gamma$ radiation.

	The nuclear vibrations can be considered like the motion of a classical droplet.
Write down the conventional expansion of the nuclear form in spherical
harmonics
\be
R(\theta,\phi)=R_0\left(1+\beta_0+\sum_{\lambda,\mu}\beta_{\lambda,\mu}
Y_{\lambda,\mu}(\theta,\phi)\right)\;.
\label{Deq1}
\ee
Main properties of the fragments can be described by a I lowing for the
quadrupole and octupole terms. Such superposition leads to a pear-like form of the nucleus. It is
essential that the electric dipole term in this case must be included in eq. (\ref{Deq1})
to keep the centre of mass fixed \cite{D6}, the relation
$\beta_1= -0.743 \beta_2\beta_3$ following from the latter condition~\cite{D1,D7}.
The other consequence is the appearance of the polarization electric dipole moment
in the nucleus~\cite{D6}:
\be
d\equiv D/e = -\kappa \beta_2\beta_3\;. \label{Deq2}
\ee
Polarizability $\lambda$ (or $\kappa$ ) in eq. (\ref{Deq2}) can be evaluated
e.g. from formulae~\cite{D6,D7}, which agree with experiment  (see also
other refs. in \cite{D7}).

Considering the oscillations quasiclassically and taking into account the relaxation, put down
\be
\beta_i(t) = \beta_i^{(0)} \sin \omega_i t \, \exp (-\gamma_it/2)\;, \quad i=2, \,3\;.
\label{Deq3}
\ee
Then the spectral density of the radiated energy is given in the classical limit~\cite{D8}
by the following expression:
\be
d{\cal E}_\omega = \frac43 \bigl |\ddot D_\omega \bigr|^2\;,    \label{Deq4}
\ee
where $D_\omega$ is the Fourier transform of the second derivative of $D(t)$ with respect to time.
Using (\ref{Deq2}) and (\ref{Deq3}) in eq. (\ref{Deq4}), we find
\bea
\ddot D_\omega = -\kappa \beta_2^{(0)} \beta_3^{(0)}
\int_0^\infty \exp(i\omega t)\frac{d^2}{dt^2}
\sin\omega_2t \sin\omega_3t \exp(-\gamma t/2)= \nonumber \\
= i D_0 \frac{(\omega_2+\omega_3)^2}{\omega-\omega_2-\omega_3+i\frac\gamma 2}\;,
\label{Deq5}
\eea
where $\gamma=\gamma_2+\gamma_3$ is the total quenching, and
$D_0=-\kappa \beta_2^{(0)} \beta_3^{(0)}$.

Supposing  $\beta_2^{(0)}\approx\beta_3^{(0)}\approx 0.7$~\cite{D5},
we calculate by means of
formulae~\cite{D7} \mbox{$d_0\approx 5$~Fm}. Using then the LDM values for a representative
heavy fragment $^{140}$Xe,
which are \mbox{$\hbar \omega_2$ = 2.2 MeV},
\mbox{$\hbar \omega_3$ = 2.8 MeV}, and evaluating $\gamma$ from
the lifetime $\tau_{diss}=\gamma^{-1}=10^{-19}$~s,
as it is stated previously, one immediately finds by means of eq. (\ref{Deq4})
\be
N_\gamma = \int_0^\infty \frac{d{\cal E}_\omega}{\hbar \omega}
\approx 8\cdot 10^{-3}\;\mbox{fission$^{-1}$}\;.
\label{Deq6}
\ee
We conclude that this value is in qualitative agreement with
experiment~\cite{D3,D4}, taking into account the uncertainties
connected with the value of $\tau_{diss}$.
For the value supposed, $\tau_{diss}\approx 10^{-19}$~s,
the contribution of the proposed mechanism is enough to explain the
experimental value.

On the other hand, we see that this contribution is proportional to $\tau_{diss}$.
Therefore, study of non-statistical $\gamma$ rays from fission gives
direct information about dissipation in largo-amplitude collective
motion represented by postrupture oscillations in the fragments.

\end{document}